\documentstyle[epsfig,12pt]{article}
\begin{document}
\begin{titlepage}
\centerline{\large \bf Significance of Single Pion Exchange  }
\vspace{0.2cm}
\centerline{\large\bf Inelastic Final State Interaction for
$D\rightarrow VP$ Processes}

\vspace{0.8cm}

\centerline{\bf Xue-Qian Li$^{a,b}$ and Bing-Song Zou$^c$}
\vspace{0.6cm}
{\small
\centerline{a) Rutherford Appleton Laboratory, Chilton, Didcot, OX11 0QX, UK}
\centerline{b) Department of Physics, Nankai University, Tianjin 300071, China}
\centerline{c) Queen Mary and Westfield College, London E1 4NS, UK}
}
\vspace{2cm}

\begin{center}
\begin{minipage}{12cm}
{\small
\centerline{\large \bf Abstract}
\vspace{1cm}

We evaluate the contribution of the final state interaction (FSI) 
due to single pion exchange inelastic scattering for 
$D^+\rightarrow \bar {K}^{0*}\pi^+$ and
$D^+\rightarrow \bar K^0\rho^+$ processes. The effects are found to be very
significant. The hadronic matrix elements of the weak transition
are calculated in terms of the heavy quark effective theory
(HQET), so are less model-dependent and more reliable.
}

\end{minipage}
\end{center}
\vspace{2cm}

PACS number(s): 13.75.Lb, 13.20.Gd, 13.25, 14.40

\end{titlepage}
\baselineskip 24pt

\noindent{\bf I. Introduction}
\vspace{0.3cm}

To understand precisely the mechanism  governing weak transition processes
where fundamental physics and possible new physics apply, one needs 
to face a synthesis  
problem including evaluation of the hadronic matrix elements and final state
interactions (FSI). Unless one can fully understand the side effects, such 
as FSI, one can hardly extract information about new physics  correctly.
At least one needs to know the order of magnitude of FSI and determine its
significance.

Due to the success of the Standard Model, the Hamiltonian for the weak 
transition $c\rightarrow s+u+\bar d$ is well understood \cite{Gilman}, and 
thanks to the heavy quark effective theory (HQET) \cite{Isgur}, we have some
reliable ways to handle the hadronic matrix elements for the 
$b\rightarrow c$ transition. The developments of the chiral lagrangian 
\cite{Wise}
enable us to estimate matrix elements from heavy mesons to
light pseudoscalars. 
According to the recent work by Roberts and Ledroit \cite{Roberts}, the
transition matrix elements from $B$, $D$ to $K^{(*)}$, become calculable in
a unique theoretical framework.

On the other hand, to really test the theory concerning
matrix elements and HQET, one needs to calculate
FSI effects which sometimes become very significant or 
sometimes can be negligible.

In the early work, Buccella et al. \cite{Buccella} found  that the calculated
branching ratios for $D^+\rightarrow \bar K^{*0}\pi^+$ and $D^+\rightarrow
\bar K^0\rho^+$ deviated significantly from the data and even using recent
data, the difference is still obvious. We use the HQET to recalculate (see below
for details) and find that the discrepancy with data is reduced, but still
exists. This disagreement may be caused by the final state
interaction (FSI). In fact, many authors have studied the problem of 
FSI in some D and B decays \cite{Buccella,Close,Zheng,Donoghue}. 
The FSI due to s-channel resonances were found to be very important
for some final channels\cite{Buccella,Close}. However for the processes
$D^+\rightarrow \bar K^{*0}\pi^+$ and $D^+\rightarrow
\bar K^0\rho^+$, the final states have isospin I=3/2, and therefore have no
s-channel resonance FSI. Zheng \cite{Zheng} calculated the elastic FSI 
effects in $B\rightarrow DK$ caused by t-channel meson exchange, 
and obtained a very small phase,  so it does not make a substantial effect to 
the measured rate. Very recently, Donoghue et al. indicate that inelastic 
scattering may dominate the FSI \cite{Donoghue}. We concur with this.
Our previous work in $\Psi$ decays and $\bar pp$ annihilations \cite{Zou} 
indicate that
when a single$-\pi$ exchange in the t-channel can be realized, the 
corresponding mechanism would make a significant contribution to the FSI and 
would very probably dominate.

For $D^+\rightarrow \bar K^{*0}\pi^+$ and 
$D^+\rightarrow \bar K^0\rho^+$, the dominant FSI should be 
the inelastic rescattering between $\bar K^{*0}\pi^+$
and $\bar K^0\rho^+$, as shown in Fig.1. In this work we estimate this
inelastic FSI effect.
The hadronic matrix elements are evaluated in terms of the
method given in ref.\cite{Roberts} which can alleviate the model-dependence of
the calculations.

In Sec.II, we give the formulation of the transition amplitude with and
without considering the inelastic scattering of $\bar K^{*0}\pi^+
\leftrightarrow \bar K^0\rho^+$, and in Sec.III, we present our numerical
results, while the last section is devoted to our discussion and conclusion.\\

\baselineskip 23pt
\noindent {\bf II. The formulation}
\vspace{0.3cm}

In the weak interaction, isospin is not conserved. 
There are four possible VP decay modes for $D^+$, i.e., $\bar K^{*0}\pi^+$,
$\bar K\rho^+$, $K^{*+}\pi^0$ and $K^+\rho^0$; among them
$I_3(\bar K^{*0}\pi^+)=I_3(\bar K^0\rho^+)=3/2$ and 
$I_3(K^{*+}\rho^0)=I_3(K^{*+}\pi^0)=1/2$.
However, from the quark diagrams \cite{Bauer}, one finds
$D^+\rightarrow K^{*+}\pi^0$ or $D^+\rightarrow K^{+}\rho^0$ can only be 
realized via a Cabibbo doubly suppressed channel, so must be very small and
can be neglected. The FSI conserves isospin. 
The $\bar K^{*0}\pi^+$ and $\bar K^0\rho^+$
cannot rescatter into  $K^{*+}\rho^0$ and $K^{*+}\pi^0$;
while $\bar K^{*0}\pi^+\leftrightarrow \bar K^0\rho^+$ can be realized
by the t-channel single pion exchange diagrams shown in Fig.1 and can
be very important. Therefore, we
only need to consider production of $D^+\rightarrow \bar K^{*0}\pi^+$ and
$D^+\rightarrow \bar K^0\rho^+$, as well as their mutual conversion 
through inelastic scattering.
 
(i) Without the FSI.

The weak interaction hamiltonian for non-leptonic decay is given as
\begin{equation}
H_W={G_F\over \sqrt 2}V_{ud}V^*_{cs}[c_1\bar s\gamma_{\mu}(1-\gamma_5)c
\bar u\gamma^{\mu}(1-\gamma_5)d + c_2\bar s\gamma_{\mu}(1-\gamma_5)d
\bar u\gamma^{\mu}(1-\gamma_5)c]
\end{equation}
where $V_{ud}$ and $V_{cs}$ are the Cabibbo-Kobayashi-Maskawa entries. 
The reactions concerned are Cabibbo favored processes.  
The color indices are dropped out as well understood, and the coefficients
$c_1$ and $c_2$ are obtained from the renormalization group equation
\cite{Gilman}.

In the calculations, we use vacuum saturation and ignore the W-exchange
(annihilation) contributions \cite{Bauer}. The transition matrix elements
for $D^+\rightarrow \bar K^{*0}\pi^+$ and $D^+\rightarrow \bar K^{0}\rho^+$
are :

\begin{eqnarray}
\label{dog}
T_1(D^+\to \bar K^{*0}\pi^+) &=& (c_1+{1\over N_c}(1+\delta)c_2)
<\pi^+|\bar u\gamma^{\mu}(1-\gamma_5)d|0><\bar K^{*0}|\bar s\gamma_{\mu}
(1-\gamma_5)c|D^+> + \nonumber \\
&+& (c_2+{1\over N_c}(1+\delta)c_1)
<\bar K^{*0}|\bar s\gamma^{\mu}(1-\gamma_5)d|0><\pi^+|\bar u
\gamma_{\mu}(1-\gamma_5)c|D^+> \nonumber \\
&=& -(c_1+{1\over N_c}(1+\delta)c_2)f_{\pi}p_{\pi}^{\mu}Tr\{[(\xi_3+
\rlap /p_{_{K^*}}\xi_4)\epsilon_{_{K^*}}\cdot v \nonumber\\
&+& \rlap /\epsilon_{_{K^*}}^*
(\xi_5+\rlap /p_{_{K^*}}\xi_6)]\gamma_{\mu}(1-\gamma_5)\cdot {1\over 2}
\sqrt {M_D}\gamma_5(1-\rlap /v)\} \nonumber \\
&+& (c_2+{1\over N_c}(1+\delta)c_1)f_{_{K^*}}\epsilon^{\mu*}_{_{K^*}}
m_{_{K^*}}Tr\{(\xi_1+\rlap /p_{\pi}\xi_2)\gamma_5\gamma_{\mu}\cdot
{1\over 2}\sqrt{M_D}\gamma_5(1-\rlap /v)\}, \nonumber\\
& &
\end{eqnarray}
and 
\begin{eqnarray}
\label{cat}
T_2(D^+\rightarrow \bar K^{0}\rho^+) &=& (c_1+{1\over N_c}(1+\delta)c_2)
<\rho^+|\bar u\gamma^{\mu}(1-\gamma_5)d|0><\bar K^{0}|\bar s\gamma_{\mu}
(1-\gamma_5)c|D^+> + \nonumber \\
&+& (c_2+{1\over N_c}(1+\delta)c_1)
<\bar K^{0}|\bar s\gamma^{\mu}(1-\gamma_5)d|0><\rho^+|\bar u
\gamma_{\mu}
(1-\gamma_5)c|D^+>  \nonumber \\
&=& (c_1+{1\over N_c}(1+\delta)c_2)f_{\rho}m_{\rho}
\epsilon_{\rho}^{*\mu}Tr\{(\xi_1+\rlap /p_{(K)}\xi_2)\gamma_5
\gamma_{\mu}\cdot{1\over 2}\sqrt{M_D}\gamma_5(1-\rlap /v)\} \nonumber \\
&-& (c_2+{1\over N_c}(1+\delta)c_1)f_{_{K}}p_{_{K}}^{\mu}
Tr\{[(\xi_3+
\rlap /p_{\rho}\xi_4)\epsilon_{\rho}\cdot v \nonumber \\
&+& \rlap /\epsilon_{\rho}^*
(\xi_5+\rlap /p_{\rho}\xi_6)]\gamma_{\mu}(1-\gamma_5)\cdot {1\over 2}
\sqrt {M_D}\gamma_5(1-\rlap /v)\}, 
\end{eqnarray}
where $v$ is the four-velocity of $D^+$ as $p_D^{\mu}=M_Dv^{\mu}$ and 
$\xi_i$ (i=1,...,6) are functions of momenta given in ref.\cite{Roberts};
$\delta$ is a non-factorization term and cannot be evaluated in perturbative
QCD \cite{Buras,Li}.
Recently, Sharma et al.\cite{Sharma} investigated the non-factorization
effects in $D\rightarrow PV$ decays.
Blok and Shifman\cite{Blok} give a more theoretical estimation
of the factor as 
\begin{equation}
\delta=-N_c{xm_{\sigma H}^2\over 4\pi^2f_{\pi}^2},
\end{equation}
where $x\sim 1$ and $m_{\sigma H}$ is a numerical factor. 
Admittedly\cite{Blok},
one cannot take the number as accurate, so we keep it 
as a parameter in the region of -0.5$\sim$ -1.0. In fact, in our calculations,
we take $\delta=-0.5$. Since we are mainly discussing the significance of
FSI, the choice of $\delta$ does not influence our qualitative conclusion
at all.

(ii) With the final state interaction.

Here, as discussed in our previous work \cite{Zou} on the FSI, 
one only needs to
calculate the absorptive part of the diagram. According to the
Cutkosky rule, for getting the
absorptive part of the loop shown in Fig.1, there are two ways to make cuts, 
i.e. (1) let the V and P be on mass-shell
while retaining the t-channel pion off-shell; (2) let the pion and P be on shell
while leaving V off-shell. The second way refers to a three-body decay
process and numerical computation for similar triangle diagrams in 
$\bar pp$ annihilation\cite{Locher} shows that it is much smaller than (1), so
we omit the possibility in our later formulation.

(a) For $D^+\rightarrow \bar K^{0*}\pi^+\rightarrow \bar K^0\rho^+$.

In the CM frame of $D^+$ where $v=(1, \vec 0)$, the
calculation can be greatly simplified. To obtain the absorptive part of the 
loop, for example $T_3$, one can just start from eq.(\ref{dog}), replace
$\epsilon_{K^*}^{\mu}$ by ${1\over 2}(2\pi)^2\delta(p_1^2-m_1^2)\delta(p_3^2
-m_3^2)(-g_{\mu\mu'}+{p_{1\mu}p_{1\mu'}\over m_1^2})$ and add the effective
vertices of the strong interaction as well as the propagator of the t-channel
pion.

\begin{eqnarray}
T_3 &=& \int{d^4p_1\over (2\pi)^4}(p_2-p_{\pi^0})^{\mu}\epsilon^*_{\rho}
\cdot (p_3-p_{\pi^0}){F(p^2_{\pi^0})\over p_{\pi^0}^2-m_{\pi^0}^2}
{1\over 2}(2\pi)^2\delta(p_1^2-m_1^2)\delta(p_3^2-m_3^2)g_{\rho\pi\pi}
g_{_{K^*K\pi}}\nonumber \\
&\times & \{-2(c_1+{1\over N_c}(1+\delta)c_2)f_{\pi}\sqrt{M_D}[(\xi_3E_3-
{1\over 2}(\xi_4-\xi_6)(M_D^2-m_1^2-m_3^2))(-g_{0\mu}+{p_{10}p_{1\mu}\over 
m_1^2})\nonumber \\
&-& (\xi_5+\xi_6E_1)p_3^{\mu'}(-g_{\mu\mu'}+{p_{1\mu}p_{1\mu'}\over m_1^2})
+i\xi_6\epsilon^{ijk}p_{1j}p_{3k}(-g_{i\mu}+{p_{1i}p_{1\mu}\over m_1^2})]
\nonumber \\
&+& [2(c_2+{1\over N_c}(1+\delta)c_1)f_{_{K^*}}m_1\sqrt{M_D}(\xi_1(-g_{0\mu}
+{p_{10}p_{1\mu}\over m_1^2})-\xi_2p_3^{\mu'}(-g_{\mu\mu'}+{p_{1\mu}p_{1\mu'}
\over m_1^2}))]\}.\nonumber\\
& &
\end{eqnarray}

(b) For $D^+\rightarrow \bar K^0\rho^+\rightarrow\bar K^{*0}\pi^+$.

The amplitude $T_4$ has a similar form as $T_3$ with some changes, and can
be easy to obtain from \ref{cat}.  Saving space, we do not 
give the explicit expression here.

In the expressions, for simplicity of bookkeeping, we set 
$ p_1\equiv p_{_{K^*}}, \; p_2\equiv p_{_{K}}, \; p_3\equiv p_{\pi^+},
\; p_4\equiv p_{\rho}$
and
$ m_1\equiv m_{_{K^*}}, \; m_2\equiv m_{_{K}},\; m_3\equiv m_{\pi^+},
\; m_4\equiv m_{\rho}.$

$F(p^2_{\pi^0})$ is an off-shell form factor for the vertices $\rho\pi\pi$
and $\bar K^{*0}\bar K^0\pi$. Because we use the 
experimental data where all the three 
particles are real and on mass-shell to fix the effective coupling at the
vertices, in the case of the pion being off-shell, a compensation form factor
is needed and it is
\begin{equation}
\label{tiger}
F(p_{\pi^0}^2)=({\Lambda^2-m_{\pi^0}^2 \over p_{\pi^0}^2-m_{\pi^0}^2})^2,
\end{equation}
with $\Lambda$ in the range of $1.2\sim 2.0$ GeV. 

From the above equations and employing the 
helicity-coupling amplitude formalism given by Chung \cite{Chung}, the whole
calculation is straightforward though tedious, 
and a direct comparison at the amplitude
level is feasible. We present our numerical results in the next section.\\

\noindent{\bf III. The numerical results.}

\vspace{0.3cm}

The two strong coupling constants $g_{\rho\pi\pi}$ and $g_{_{K^*K\pi}}$
can be obtained from the $\rho$ and $K^*$ decay width, respectively.
From the newest PDG data \cite{Data}, we have $g_{\rho\pi\pi}=6.1$
and $g_{_{K^*K\pi}}=5.8$. The values of $c_1$ and $c_2$ are taken from
Ref.\cite{Sharma}, i.e., $c_1=1.26$ and $c_2=-0.51$.

As aforementioned, we take the non-factorization factor $\delta$ to be -0.5.
The $\xi_i$'s (i=1,...,6) have simple Gaussian forms or polynomials. Their
explicit forms and parameters can be found in ref.\cite{Roberts}. There are 
three sets of parameters for $a_i$ and $b_i$ (notation in ref.\cite{Roberts}),
which look very different. We substitute all the three sets into our 
expressions to carry out the calculations and compare their results.

(i) First, we calculate the decay rate without taking into account the FSI, i.e.
we only use eqs. (\ref{dog}) and (\ref{cat}). The results are listed in Table I.
\\

\begin{center}
\begin{tabular}{|c|c|c|c|c|}
\hline
& Fit 1 & Fit 2 & Fit 3 & Exp. \\
\hline
$\Gamma(D^+\rightarrow\bar K^0\rho^+)\; (\times 10^{-13}$ GeV) & 1.0 & 1.0 &
0.65 & $0.41\pm 0.15$ \\
\hline 
$\Gamma(D^+\rightarrow \bar K^{*0}\pi^+)\; (\times 10^{-13}$ GeV) & 0.09 &
0.049 & 0.079 & $0.124\pm 0.025$ \\
\hline
\end{tabular}
\vspace{0.2cm}

 Table I. Results without considering FSI and with three sets of parameters
from \cite{Roberts}.

\vspace{0.2cm}
\end{center}

Here Fit1, Fit2 and Fit3 correspond to three different sets of parameters 
for the functions $\xi_i$ (i=1,...,6) from Roberts et al.\cite{Roberts}
fitting $D\rightarrow K^{(*)}l\nu$; $D\rightarrow K^{(*)}
l\nu,\; B\rightarrow K^{(*)}J/\psi$ and $D\rightarrow K^{(*)}l\nu,\;
B\rightarrow KJ/\psi,\; B\rightarrow K\psi'$, respectively. 
One notices that even though the parameters in the three fits are quite 
different, the results are rather close
to each other. These results are still qualitatively consistent with that 
obtained by authors of ref.\cite{Buccella} without the FSI, namely the 
calculated rate for $D^+\rightarrow \bar K^{*0}\pi^+$ is lower than the
experimental value by 1.4$\sim$1.6 times, while for $D^+\rightarrow\bar K^0
\rho^+$, the calculated number is $1.6\sim 2.5$ times larger than data.

(ii) The FSI contribution.

Since we only consider the absorptive part of the loop, we can get the 
imaginary part of the FSI amplitude.
The dispersive part can be obtained by the dispersive relation with some 
cut-off parameters and  should be of the same order as the absorptive
one \cite{Locher}. The absorptive part of the amplitude gives a lower
bound on the effect of FSI.

For a clean comparison and to avoid ambiguity, we use helicity amplitudes. In
the helicity picture, for the CM frame of $D^+$ all momenta of the outgoing
$P$ and $V$ are aligned along the $\hat z$ axis (or oppositely). 
Even though the mesons in the loop are real, their momenta 
can deviate from the $\hat z$ axis by an angle $\theta$.
So in this scenario, for a reaction $D\rightarrow PV$, only $\epsilon^*(0)$
polarization of the final outgoing vector meson contributes. In Table II, 
we present the numerical value for the absorptive part of the FSI
loop with $\Lambda=1.6 GeV$, and as a comparison also the helicity amplitude 
without FSI evaluated in the same theory.\\

\begin{center}
\begin{tabular}{|l|c|c|c|}
\hline
 & Fit1 & Fit2 & Fit3\\
\hline
$T(D^+\rightarrow\bar K^{0}\rho^+\rightarrow\bar K^{*0}\pi^+)$ & -i0.220
& -i0.223 & -i0.178\\
\hline
$T(D^+\rightarrow\bar K^{*0}\pi^+\rightarrow\bar K^{0}\rho^+)$ & -i0.071 &
-i0.051 & -i0.062\\
\hline
$T(D^+\rightarrow\bar K^{0}\rho^+)$ & 0.52 & 0.52 & 0.42\\
\hline
$T(D^+\rightarrow\bar K^{*0}\pi^+)$ & 0.15 & 0.11 & 0.14\\
\hline
\end{tabular}
\vspace{0.2cm}

Table II. 

\end{center}

In the table, we only keep the relative values of the helicity amplitudes 
with and without the FSI, while dropping out any common factor such as
$(G_F/\sqrt 2)V_{cs}V^*_{ud}$ etc. The values given in Table II are calculated
with $\Lambda=1.6$ GeV. In fact, our numerical computation show that
as $\Lambda$ varies from 1.2 to 2.0 GeV, the corresponding results 
in the first two rows of Table II can change by a factor 2. As one takes
a more restricted region for $\Lambda$, the results are not very sensitive
to the $\Lambda$ value.

Obviously, the FSI effect of $T(D^+\rightarrow\bar K^{0}\rho^+\rightarrow\bar 
K^{*0}\pi^+)$ is stronger than that of $T(D^+\rightarrow\bar K^{*0}\pi^+
\rightarrow\bar K^{0}\rho^+)$; we will discuss the results below.\\

\noindent{\bf IV. Conclusion and discussion}

Above, we calculate the decay width of $D^+\rightarrow\bar K^{*0}\pi^+$ and
$D^+\rightarrow\bar K^{0}\rho^+$ in terms of the HQET and consider the
contribution of the t-channel single pion exchange to the inelastic scattering
$\bar K^{*0}\pi^+\leftrightarrow\bar K^{0}\rho^+$.

From Table I, we notice that the directly calculated value without considering
theFSI for   $D^+\rightarrow\bar K^{*0}\pi^+$
is lower than the experimental value by about 1.5 times while for
$D^+\rightarrow\bar K^{0}\rho^+$ it is 1.6$\sim$2.5 times larger.
In our calculations, three different sets of parameters from   
Roberts et al.\cite{Roberts} are used for the functions $\xi_i$ (i=1,...,6) 
in the Gaussian forms (or polynomials).
The three sets of parameters are obtained by fitting 
(1) $D\rightarrow K^{(*)}l\nu$; (2) $D\rightarrow K^{(*)}
l\nu,\; B\rightarrow K^{(*)}J/\psi$; and (3) $D\rightarrow K^{(*)}l\nu,\;
B\rightarrow KJ/\psi,\; B\rightarrow K\psi'$, respectively.
Obviously, the third set of parameters fits more sets of data, and happens to
give the best fit (Fit3) to the data in our cases. 
But there is still a discrepancy.

Table II shows that the FSI effects are of similar order of magnitude 
to the direct production rates. Especially, the amplitude 
$T(D^+\rightarrow\bar K^{0}\rho^+\to\bar K^{*0}\pi^+)$
is quite large. This will reduce the rate for the $\bar K^{0}\rho^+$ final
state and raise the rate for the $K^{*0}\pi^+$ final state; therefore it will
bring the theoretical results into better agreement with the experimental data.

In this work we are not trying to fit the data numerically.
We do not introduce any free parameter. 
We just check what is the theoretical prediction for the processes
$D^+\rightarrow\bar K^{*0}\pi^+$ and
$D^+\rightarrow\bar K^{0}\rho^+$, based on the more reliable and widely 
accepted theoretical framework\cite{Roberts} without considering the FSI.
We find that the calculated values without considering the FSI obviously
deviate from the experimental data. Then we investigate the magnitude of the
FSI effects due to single pion exchange inelastic scattering with the 
unitarity approximation which has been testified in many practical
processes and proved to be reasonable \cite{Zou,Locher}. Our results indicate
that the single pion exchange process $\bar K^0\rho^+\leftrightarrow 
\bar K^{*0}\pi^+$ has significant effects for the processes $D^+\rightarrow
\bar K^0\rho^+$ and $D^+\rightarrow \bar K^{*0}\pi^+$, and may be the reason
for the discrepancy  between the experimental data and earlier theoretical
predictions.

We cannot give more accurate results, since the dispersive part of the
inelastic scattering amplitude is very model-dependent, so in this work we 
are not going to adjust the cut-off parameters for fitting data to cause
some mess and uncertainty at this stage.

Our discovery of the significance of the single pion exchange inelastic 
scattering in the FSI may have important applications to many other D and B
decays. For example, for $B\rightarrow DK$, Zheng \cite{Zheng} concluded that
the FSI from elastic scattering is negligible. It is noted that for the
elastic scattering the lightest exchanged mesons are $\sigma$ or $\rho$,
while the inelastic scattering $D^*K^*\leftrightarrow DK$ can be realized 
by exchanging pions and may give more significant contributions. The 
single pion exchange inelastic  FSI may play an important role in the channels 
which are relevant to evaluating CP violation in some channels and 
precise measurement of the CKM matrix entries. 
The investigation of those effects is in progress.\\

\noindent{\bf Acknowledgments:} 
We thank David Bugg and Chao-Hsi Chang for discussions and comments.
One of us (Li) would like to thank the Rutherford Appleton Laboratory for
its hospitality; the main part of the work is accomplished during the period
of his visit at the lab. 
The work is partially supported by the National Natural Science
Foundation of China.

\newpage

\begin{figure}[htbp]
\begin{center}\hspace*{-0.cm}
\epsfysize=15.0cm
\epsffile{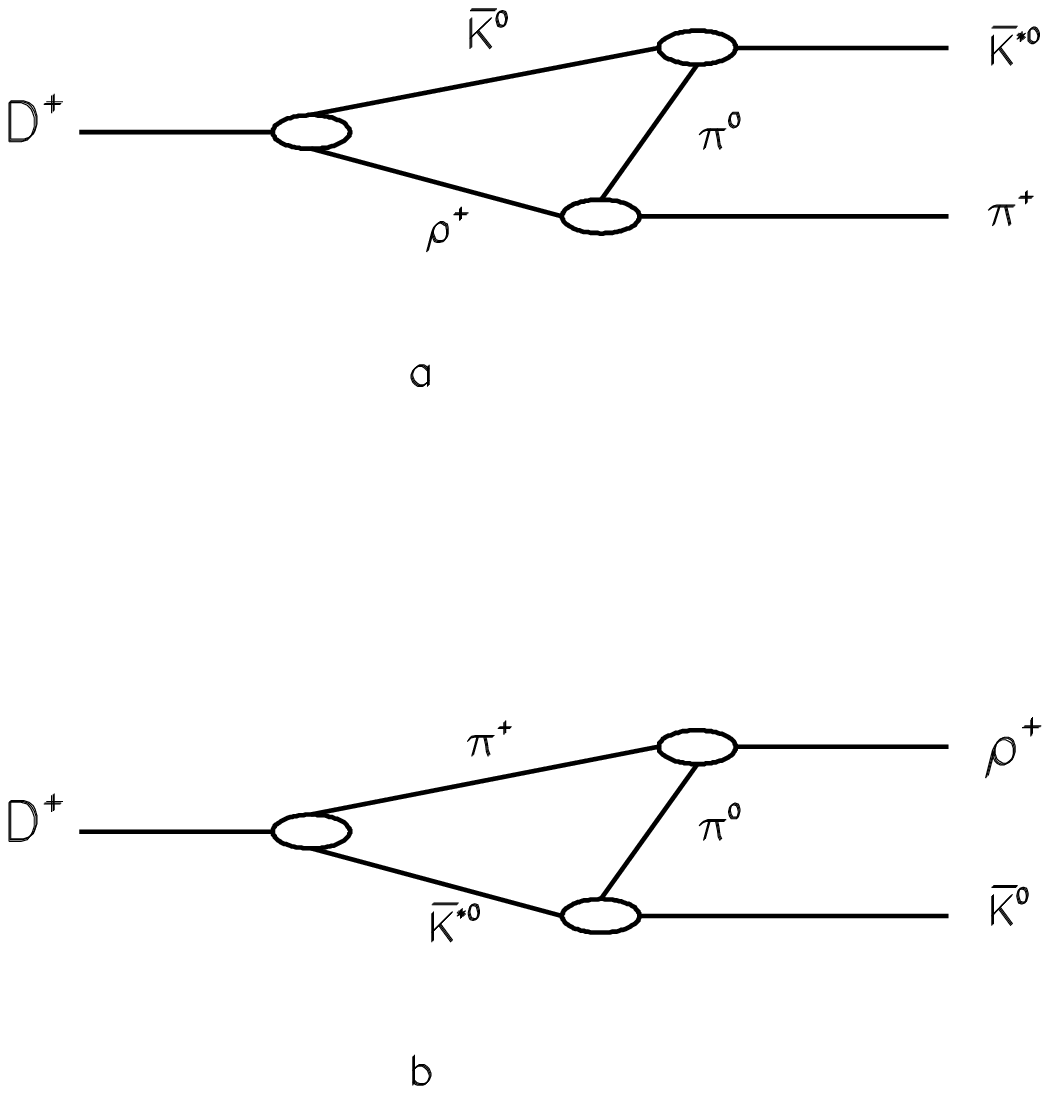}
\end{center}
\caption{The single pion exchange inelastic FSI loop for 
$D^+\to\bar K^{*0}\pi^+$ and $D^+\to\rho^+\bar K^0$
}
\label{fig:ampl}
\end{figure}
\end{document}